# Axis of Zodiacal light Near Tropic Cancer

**Nawar, S., Tadross, A.L., Mikhail, J.S., Morcos, A.B., Alhassan I. Ibrahim.**

**Abstract:** The axis of Zodiacal lights have been obtained in blue and yellow colors using photoelectric observations of Zodiacal light. These observations have been carried out at Abu-Simbel Site of Egypt in October 1975. The observational site ($\Phi=22^o$) lies too near to the Tropic Cancer, at which the axis of the Zodiacal light cone perpendiculars to the horizon. The results show that the plane of the Zodiacal light is inclined to the normal by $1.59^o$ in blue color and $1.18^o$ in yellow color. This means that there is a slight variation in Zodiacal light's axis with wavelength, and the axis almost coincide with the ecliptic. The present results for blue color can be considered as the first one in the world near the Tropic of Cancer.

## 1. INRODUCTION.

The photometric axis of the Zodiacal lights (ZL), is defined as the locus of the points of maximum brightness. The earlier observations of the ZL, carried out by Fath (1909) showed that, the maximum light displaced 1.7˚ east of the vertical circle passing through the sun. Those observations showed also the tendency of the displacement to increase with time. Hoffmeister (1940) concluded from several hundreds of visual observations that at elongation angles $\varepsilon = 25^o$ to 90˚ from the sun, the axis of symmetry corresponds to the ecliptic and the orbital planes of Venus, lies at heliocentric distances of 1.0 and 0.4 AU respectively.

Photographic observations, were taken by Blackwell & Ingham (1961) and Donitch (1953,1955), led them to conclude that, the plane of symmetry is the invariable plane of the solar system ($\Omega =107^\circ$, $i =1.6^\circ$). From the photoelectric measurements, Regner (1955) and Divari & Asaad (1960) have found that, the axis of symmetry lies very close to the ecliptic plane.

Photoelectric observations of the ZL at $\varepsilon = 90^\circ$ carried out by Huruhata (1965) and Saito & Huruhata (1967), supports the suggestion that, the symmetry plane lies close to the orbital plane of Venus (ascending node $\Omega = 76^\circ$, inclination to the ecliptic $i=3.4^\circ$) or the solar equatorial plane ($\Omega =73^\circ$, $i =7.3^\circ$).

Leinert et al (1974) reported that, the photometric axis has inclination angles of -3.8˚, - 4.1˚ and - 3.3˚ west of the sun, and 2.8˚, 2.7˚ and 2.0˚ east of the sun, at elongation angles 15˚, 21˚ and 30˚ respectively.

From balloon observation Macqueen (1968) has found that, the photometric axis makes an inclination angle of 1°-2˚ south of the ecliptic. While Macqueen et.al (1973) have found from Apollo 16 lunar orbit observations that, the photometric axis is inclined by $i =3^\circ\pm1^\circ$ north of the ecliptic, which does not agree with Macqueen (1968). Lillie (1968) used rocket observations to propose that a plane of symmetry at $\varepsilon =130^\circ$-180˚ east and west of the sun, inclined by 5˚ below the ecliptic with ascending node $\Omega =70^\circ$, which is close to the invariable plane. The Helios observations (Leinet et al. (1981)) yielded $\Omega=87^\circ$ and

i=3.0° for the symmetry plane of interplanetary dust. Vrtilek and Hauser (1995) analyzed IR data of the brightness variations at ecliptic poles, to estimate the Zodiacal light, symmetry plane to be at (i =1.7°-1.8°) and Ω =76°-77°). However the values (i=1.5° w, Ω =41°), were derived on the bases of the variety of latitude peak brightness.

From the Isophote contour of inner ZL in the region of solar elongation ε = 1° to ε = 22°, Cooper et al. (1996) have found that, there is at least 2° difference between the ecliptic plane and the symmetry planes. James et al. (1997), from CCD observations of the ZL, have found that, the axis of the line of the maximum intensity is not only of significantly south the ecliptic but also of the symmetry axis of the invariable plane.

## 2. OBSERVATIONS:

Observations of Zodiacal light have been carried out at Abu-Simbel Site. The site lies near the Tropic of Cancer ($\Phi=22^o$). The main purpose of these observations, is to study the spectral photometry of Zodiacal light and the position of its axis. The observations have been carried out in October 1975 using semi- automatic photoelectric scanner established mainly for observing Zodiacal light and other different components of the night sky. Two wideband glass filters have been used for observations having effective wavelength 4410 and 5510 for blue and yellow colors respectively. To obtain the point of maximum brightness, the observed brightness of Zodiacal light, has been corrected to:
   a) Atmospheric extinction, b) Integrated starlight and c) Air glow.

The ecliptic coordinates (Ɛ, β) of the points of maximum brightness of the Zodiacal light have been calculated..

## 3. RESULTS AND DISCUSSIOIN

We are going to use two methods to obtain the inclination of the axis of Zodiacal light. In the first method the ecliptic coordinates (Ɛ, β) of the points of maximum brightness, for different dates and colors, have been calculated and tabulated in Table 1 (a,b) . The average values of ecliptic latitude have been obtained and written at the end of Table(1) for each color. These average values give the position of the axis of ZL. It can be seen from Table (2) that, there is no change in Zodiacal light axis with wavelength. It is noticed that there is a slight change from night to night.

The second method has been used for the first time, and in which the results of Table (1) are drawn as in Figures (1, 2, and 3). These figures give the location of the points of maximum brightness for different dates in blue and yellow colors. The best fit line connects the points of each figure from Figures (1, 2, and 3) has been drawn using least square method. Using the slope of each line, the angle of inclination of the Zodiacal light axis has been obtained and the results are given in Table (3) for each color and date. It can be seen from Table (3) that there is a slight difference in the angle from day to day and from color to color. Table (3) indicates that, the axis of Zodiacal light in 22/23 inclined to the ecliptic by o.27 degree for both blue and yellow colors. While for the 21/22 date the axis inclined by 2.12 and 2.19 degrees for blue and yellow colors respectively. This means that during the same night the difference in the inclination of the axis of Zodiacal light between blue and yellow no more than 0.07 degrees. This result shows that the axis of Zodiacal light changes from night to night, while there no change with wavelength. The average value of the inclination of the axis is given at the end of Table (3).

### Table (1a)
**The coordinates of points of maximum brightness for blue color at different dates.**

| 19/20 | | 20/21 | | 22/23 | |
|---|---|---|---|---|---|
| ε | β | ε | B | ε | β |
| 58 | 3 | 57 | 3 | 61 | 3 |
| 74 | 0 | 64 | 1 | 70 | 0 |
| 44 | 1 | 73 | 3 | 78 | -1 |
| 44 | 3 | 49 | 1 | 86 | 3 |
| 52 | 3 | 57 | 2 | 96 | 5 |
| 59 | 2 | 64 | 1 | 46 | 0 |
| 67 | 3 | 74 | 1 | 56 | 1 |
| 36 | 2 | 83 | 1 | 63 | 2 |
| 43 | 3.5 | 89 | -1 | 71 | -2 |
| 52 | 1 | 36 | 4 | 44 | 2 |
| 70 | 5 | 45 | 3,5 | 53 | -.5 |
|  |  | 53 | 4 | 30 | -.5 |
|  |  | 60 | 3 | 39 | 1.5 |
|  |  | 29 | 3 |  |  |
|  | 2.41 |  | 2.03 |  | 0.96 |

### Table (1b)
**The coordinates of points of maximum brightness for yellow color at different dates.**

| 19/20 | | 20/21 | | 22/23 | |
|---|---|---|---|---|---|
| ε | β | ε | β | ε | β |
| 58 | 3 | 57 | 3 | 61 | 3 |
| 74 | 0 | 64 | 1 | 74 | 2 |
| 44 | 1 | 73 | 3 | 78 | -1 |
| 44 | 3 | 49 | 1 | 86 | 3 |
| 52 | 1 | 57 | 2 | 96 | 5 |
| 59 | 2 | 64 | 1 | 46 | 0 |
| 67 | 5 | 74 | 1 | 56 | -3 |
| 36 | 2 | 83 | 1 | 63 | 2 |
| 43 | 3.5 | 89 | -1 | 71 | -3 |
| 52 | 1 | 36 | 4 | 44 | 2 |
| 70 | 5 | 45 | 3.5 | 53 | -.5 |
|  |  | 53 | 4 | 30 | -.5 |
|  |  | 29 | 3 | 39 | 1.5 |
|  | 2.41 |  | 2.04 |  | 0.81 |

### Table (2)
**Angle of inclination of Zodiacal light Using the first method**

| Date | Blue | Yellow |
|---|---|---|
| 19/20 | 2.41 | 2.4 |
| 20/21 | 2.03 | 2.04 |
| 22/23 | 0.96 | 0.81 |
| Average | 1.8 | 1.75 |

### Table (3)
**Angle of inclination of Zodiacal light using the second method.**

| Date | Blue | Yellow |
|---|---|---|
| 19/20 | 2.63 | 1.09 |
| 20/21 | 2.12 | 2.19 |
| 22/23 | 0.27 | 0.27 |
| Average | 1.59 | 1.18 |

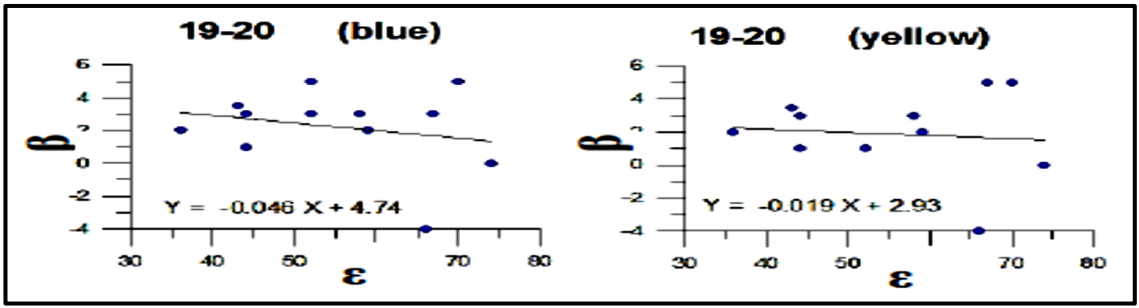

**Figure (1): The position of the points of maximum brightness for the date 19-20/10.**

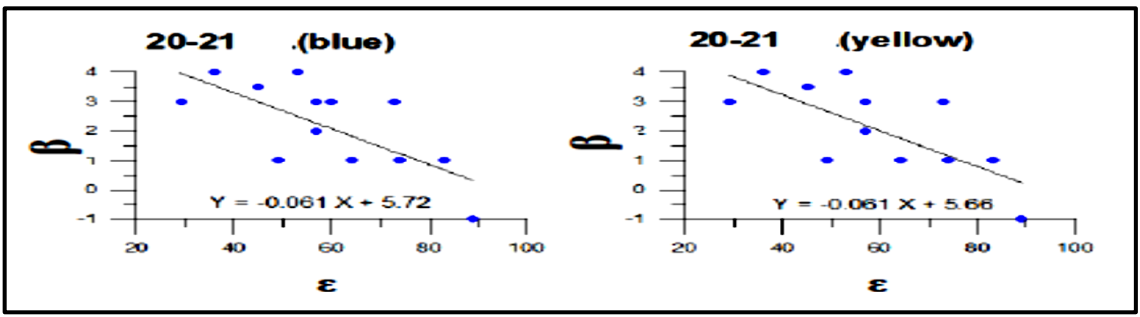

**Figure (2): The position of the points of maximum brightness for the date 20-21/10.**

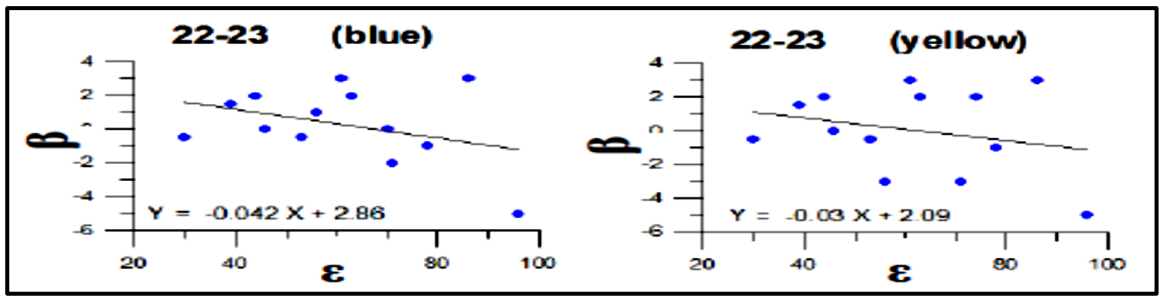

**Figure (3): The position of the points of maximum brightness for the date 22-23/10.**

The present results have been compared with that obtained near the Tropic of cancer by Divari and Assad (1960) at Daraw (Egypt). It has been found that the difference between our results and that of Divari and Asaad about 0.08 degrees. It can be seen from Table (4) That the maximum inclination of Zodiacal light axis is 7.3 degrees obtained by Huruhata (1974), while the minimum value is 1 degree obtained by Divari and Assad (1960) and Games by (1997).

## Table (4)

Comparison between the present results and that obtained by other investigators.

| Author | Year | Inclination of Axis | Remarks |
|---|---|---|---|
| **Fath** | **1909** | **1.7** | Visual |
| **Hofmiester** | **1940** | **2.0** | Photometric |
| **Blackwell** | **1961** | **1.6** | Photographic |
| **Divari and Assad** | **1960** | **1.0** | Photoelectric |
| **Macqueen** | **1968** | **2.0** | Balloon |
| **Laille** | **1968** | **5.0** | Photoelectric |
| **Hurhate** | **1974** | **7.3** | Photoelectric |
| **Leinert** | **1981** | **-3.3** | Photoelectric |
| **Leinert** | **1981** | **2 .0** | Photoelectric |
| **Hauser** | **1995** | **1.8** | Photoelectric |
| **Cooper** | **1996** | **2.0** | Photoelectric |
| **Games** | **1997** | **1.0** | CCD |
| **Present work** | **2013** | **1.2** | Photoelectric |